\documentclass{pasj02}

\usepackage[whole]{bxcjkjatype}
\usepackage{natbib}
\usepackage{url}
\usepackage[switch,mathlines]{lineno}

\def\arcdeg{$^\circ$}

\newcommand{\nar}{New Astron. Rev.}

\Received{2026/03/07}
\Accepted{2026/05/10}
\Published{yyyy/mm/dd}

\begin{document} 

\title{
Axisymmetric Jeans modelling systematically overestimates the circular speed in the inner Milky Way
}

\author{
Tomoki \textsc{Yamaguchi}\altaffilmark{1,2}
\email{tomoki.yamaguchi.astro@gmail.com}
}
\author{
Junichi \textsc{Baba}\altaffilmark{1,3} \orcid{0000-0002-2154-8740}
\email{babajn2000@gmail.com; junichi.baba@sci.kagoshima-u.ac.jp}
}

\altaffiltext{1}{Amanogawa Galaxy Astronomy Research Center, Graduate School of Science and Engineering, Kagoshima University, 1-21-35 Korimoto, Kagoshima 890-0065, Japan.}
\altaffiltext{2}{Graduate School of Science, Nagoya University, Furo-cho, Chikusa-ku, Nagoya 464–8602, Japan.}
\altaffiltext{3}{Division of Science, National Astronomical Observatory of Japan, Mitaka, Tokyo 181-8588, Japan.}


\maketitle

\begin{abstract}
We quantify systematic biases in rotation curves inferred from steady, axisymmetric Jeans modelling when the underlying stellar velocity field is non-axisymmetric.
Using a high-resolution $N$-body/hydrodynamic simulation of an isolated Milky Way--like disk galaxy, we construct mock stellar-kinematic measurements for two observer azimuths relative to the bar. One observer is placed at a Solar-like viewing angle of $25^\circ$ from the bar major axis, and the other at $115^\circ$.
For each configuration, we analyse multiple snapshots and compare the Jeans-inferred circular-speed curve $V_{\rm c,Jeans}(R)$ with a reference axisymmetric circular-speed curve $V_{\rm c,axi}(R)$ defined from the azimuthally averaged ($m=0$) component of the gravitational field.
The Jeans analysis is performed in a wedge-shaped mock observational volume that mimics limited sky coverage.
For the $25^\circ$ configuration, the mean azimuthal streaming is typically higher than the azimuthally averaged expectation by $\approx 10$--$15~\mathrm{km\,s^{-1}}$, which leads to an average overestimate of the axisymmetrically defined circular speed by $\approx 4\%$ (corresponding to $\approx 10~\mathrm{km\,s^{-1}}$) in the inner disk.
Across snapshots, the mean offset corresponds to a $\sim 1.5$--$2\sigma$ systematic deviation of $V_{\rm c,Jeans}$ from $V_{\rm c,axi}$.
For the $115^\circ$ configuration, the bias reverses sign and $V_{\rm c,Jeans}$ tends to underestimate $V_{\rm c,axi}$.
As a scaling under the usual spherical approximation, for the $25^\circ$ configuration a $\approx 4\%$ bias in $V_{\rm c}$ corresponds to an $\approx 8\%$ bias in the enclosed dynamical mass at fixed radius.
These results imply that steady, axisymmetric Jeans modelling of Milky Way stellar kinematics can overestimate the axisymmetrically defined circular-speed curve at the percent level unless non-axisymmetric streaming is modelled explicitly or the bias is included in the error budget.
\end{abstract}


\section{Introduction}
\label{sec:intro}

Constraining the dynamical mass distribution in disk galaxies, including the Milky Way, is essential for understanding the relative contributions of baryons and dark matter \citep{Courteau+2014}.
One of the most direct probes is the ``circular-speed curve'' (often referred to as the rotation curve)\footnote{
Throughout this paper, we define the \emph{circular speed} from the axisymmetric (azimuthal Fourier $m=0$) component of the gravitational potential as
$V_{\rm c}^2(R)\equiv R\,\partial\Phi_0(R,0)/\partial R$ \citep[][]{BinneyTremaine2008}.
This is the speed that balances centrifugal force with the \emph{axisymmetric} gravitational field in the disk mid-plane.
}, which traces the radial dependence of the gravitational field in the disk plane.
The shape and normalization of the rotation curve encode the radial gravitational field in the disk plane, and hence the underlying mass distribution \citep[][]{vanAlbada+1985}. It constrains the contributions of the stellar and gas disks in the inner Galaxy. It also sets the local dark-matter density and halo profile on larger scales \citep{deSalasWidmark2021,HuntVasiliev2025}.

Large astrometric and spectroscopic surveys, especially \textit{Gaia} \citep{GaiaCollaboration2016,Perryman2026}, have enabled high-precision determinations of the Milky Way circular-speed curve from stellar kinematics over a wide radial range \citep{Eilers+2019,Zhou+2023,Jiao+2023,Ou+2024,Poder+2023,Feng+2026}.
A widely used approach assumes an axisymmetric, steady-state system.
It then applies the Jeans equations to observed tracer density and velocity moments.
For example, \citet{Eilers+2019} derived a precise circular-speed curve over $R\simeq 5$--$25~\mathrm{kpc}$ from red-giant tracers using the radial Jeans equation with an asymmetric-drift correction.
Recent \textit{Gaia} DR3 analyses have adopted closely related frameworks.
They have examined systematic effects, including distance systematics, tracer selection, and departures from equilibrium \citep[][]{Jiao+2023,Poder+2023,Koop+2024}.
Such departures can be particularly important in the outer disk.
For instance, \citet{Koop+2024} argued that time-dependent perturbations can produce systematic offsets between Jeans-based estimates and the underlying circular-speed curve at large radii.
Despite these caveats, these Jeans-based determinations show very good mutual consistency at small radii ($5 \lesssim R \lesssim 10~\mathrm{kpc}$), with differences at the level of $\sim 5~\mathrm{km\,s^{-1}}$ \citep[see Fig.~17 of][]{HuntVasiliev2025}. However, this mutual consistency does not by itself ensure accuracy with respect to the axisymmetrically defined circular-speed curve.

The Milky Way is not axisymmetric.
Observations indicate a central bar extending to $\sim 5~\mathrm{kpc}$ and inclined by $\phi_{\rm bar}\sim 25^\circ$ to the Sun--Galactic-center line \citep{WeggGerhard2013,Portail+2017}, as well as prominent spiral structure traced by young stars, gas, star-forming regions, and old stars \citep[e.g.][]{HouHan2014,Reid+2019,Poggio+2021,Drimmel+2025,Miyachi+2019,Lin+2022}.
Both the bar and spiral arms drive in-plane non-circular motions that produce azimuth-dependent mean velocities, as seen in stellar kinematics in and beyond the solar neighbourhood \citep[e.g.][]{Gaia-Katz+2018,Eilers+2020} and around spiral arms \citep[e.g.][]{Baba+2018,Funakoshi+2024}.
Non-axisymmetric perturbations can also generate characteristic kinematic substructure in the disk.
For example, \textit{Gaia} data show prominent ridges in the $R$--$v_\phi$ plane \citep[e.g.][]{Kawata+2018,Antoja+2018,Ramos+2018}.
Such ridges can be interpreted as kinematic responses to bar--spiral perturbations and other time-dependent forcing \citep[e.g.][]{Hunt+2018,Fragkoudi+2019,Martinez-Medina+2019,Asano+2020}.
The same non-axisymmetric forcing also drives non-circular motions in the gaseous component.
As a result, rotation curves inferred from gas kinematics via the tangent-point method can be biased, as demonstrated with hydrodynamic Milky Way models \citep{Chemin+2015,Baba2025b,Davis+2026}.

These facts motivate a careful assessment of how much an axisymmetric Jeans analysis can deviate from the underlying axisymmetric circular speed.
Many Jeans-based studies have probed non-axisymmetry through azimuthal robustness tests.
For example, \citet{Eilers+2019} excluded $R\lesssim 5~\mathrm{kpc}$ to reduce the impact of the bar.
They then split their $60^\circ$ wedge into two disjoint $30^\circ$ wedges.
Near the Solar radius, they found a difference of $\sim 2.0~\mathrm{km\,s^{-1}}$, which is about the $1\%$ level.
Similarly, \citet{Zhou+2023} reported agreement within $\sim 2\%$ across several azimuthal bins.
They included this in their systematic error budget. 
\citet{Feng+2026} similarly reported differences at the $3\%$ level in wedge-splitting tests.

Azimuthal-splitting tests are informative, but they mainly constrain \emph{relative} variations across the surveyed wedges.
They are therefore most sensitive to azimuthal gradients within the selected region, and in practice they often yield only few-percent differences even in recent applications \citep[e.g.][]{Feng+2026}.
If non-circular motions are coherent over the full azimuthal range accessible to the data, wedge-to-wedge differences can remain small.
This can occur even when the \emph{absolute} offset from the axisymmetrically defined circular speed is not negligible.
Related issues have been discussed for \emph{local} determinations near the Solar radius.
Using classical Cepheids, \citet{Kawata+2019} emphasized that simple axisymmetric modeling recovers a local centrifugal speed at the observer position.
This local quantity can differ from the azimuthally averaged circular speed when the radial force varies with azimuth due to the bar and spirals.
Similarly, \citet{Almannaei+2024} showed with \textit{Gaia} DR3 young stars that the Local arm can induce spatial variations in the inferred local circular speed under axisymmetric assumptions.
Together, these arguments suggest that small wedge-to-wedge differences do not necessarily imply a small absolute bias, especially in regions where non-axisymmetric streaming is coherent and strong.

In this work, we focus on $5\lesssim R\lesssim 8~\mathrm{kpc}$, just outside the bar, where bar- and spiral-driven streaming remains strong and can bias Jeans-based circular-speed estimates.
To quantify this systematic bias, we use a simulation-based experiment in which a steady, axisymmetric Jeans analysis is applied to a non-axisymmetric disk. Using an $N$-body/hydrodynamic Milky Way analogue \citep{Baba2025a}, we compare the Jeans-inferred circular-speed curve with a reference axisymmetric circular-speed curve defined from the axisymmetric (azimuthal Fourier $m=0$) component of the gravitational potential.
The remainder of the paper is organized as follows. 
Section~\ref{sec:methods} describes the simulation, the mock observational volume, and the construction of the reference axisymmetric circular-speed curve, and outlines the Jeans estimator.
Section~\ref{sec:Results} presents the bias and scatter and connects them to the Jeans term budget and the non-axisymmetric streaming field.
Section~\ref{sec:Discussion} summarizes the main findings and discusses implications for Milky Way rotation-curve measurements.

\section{Methods}
\label{sec:methods}

\subsection{Simulations}
\label{sec:method_sim}

\begin{figure*}
\begin{center}
\includegraphics[width=\linewidth]{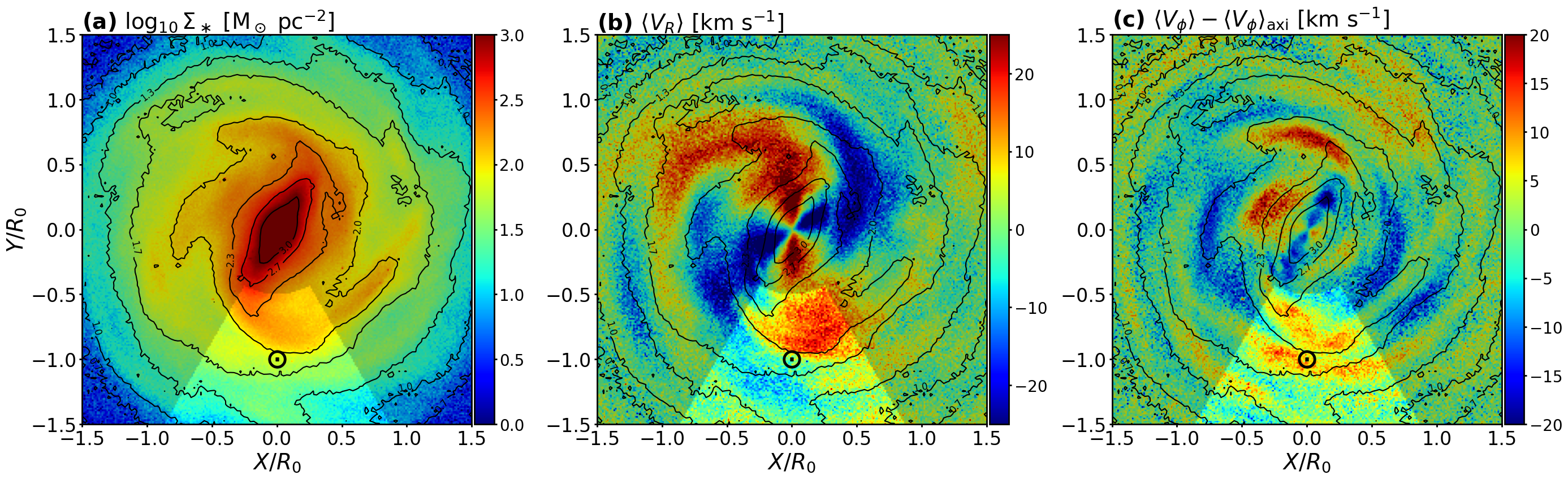}
\end{center}
\caption{
Example snapshot of the simulated stellar disk on the $(x,y)$ plane. The disk rotates clockwise in this coordinate system.
\textbf{(a)} Stellar surface-density map of $\Sigma_\ast\ [{\rm M_\odot,pc^{-2}}]$, shown on a logarithmic scale, with isodensity contours overplotted.
\textbf{(b)} Mean radial velocity field, $\langle v_R\rangle$.
\textbf{(c)} Residual mean azimuthal velocity field, $\langle v_\phi\rangle-\langle v_\phi\rangle_{\rm axi}$, where $\langle v_\phi\rangle_{\rm axi}$ is the azimuthally averaged mean azimuthal velocity.
The $\odot$ symbol marks the mock observer at $(x,y)=(0,-R_0)$.
The unshaded wedge indicates the mock observational volume used for the Jeans analysis ($|\phi|<30^\circ$ and $R/R_0>0.5$), while the shaded region shows the excluded area outside this field of view.
{\bf Alt text:}
Face-on snapshot of a barred stellar disk showing the bar and spiral structure, with corresponding non-axisymmetric patterns in the mean radial velocity and in the residual mean azimuthal velocity; the observer wedge selection is indicated.
}
\label{fig:snapshot}
\end{figure*}

We analyze the $N$-body/hydrodynamic simulation of an isolated Milky Way--like barred spiral galaxy presented in \citet{Baba2025a}.
The initial conditions were designed to resemble the Milky Way and were based on the axisymmetric mass model of \citet{McMillan2017}.
We ran the simulation with the $N$-body/smoothed particle hydrodynamics (SPH) code \texttt{ASURA-3} \citep{Saitoh+2008,Saitoh2017}.
The model includes radiative cooling and heating, star formation, and stellar feedback.
Hydrodynamics is solved with the density-independent SPH (DISPH) method \citep{SaitohMakino2013}.
A stellar bar forms spontaneously via the bar instability \citep[e.g.][]{Efstathiou+1982,Fujii+2018} at $t\simeq 1~\mathrm{Gyr}$.
By $t\simeq 1.5~\mathrm{Gyr}$, the rapid bar-growth phase has largely ended and the bar properties (e.g.\ its strength and pattern speed) evolve more slowly.
We analyze multiple snapshots in the post-bar-formation phase spanning $t=2$--$3~\mathrm{Gyr}$.

Figure~\ref{fig:snapshot} shows an example snapshot that illustrates the non-axisymmetric structure and bar-driven non-circular motions. The stellar surface-density map (panel~a) shows a well-developed bar with a semi-major axis of $\sim 3$--$4~\mathrm{kpc}$. It also shows spiral structure at larger radii. We estimate the bar pattern speed by tracking the time evolution of the phase of the $m=2$ Fourier mode of the stellar surface density within $R<3~\mathrm{kpc}$. For this snapshot, we obtain $\Omega_{\rm bar}\approx 40~\mathrm{km\,s^{-1}\,kpc^{-1}}$. The corresponding corotation radius is $R_{\rm CR}\approx 5.5~\mathrm{kpc}$. Overall, the bar properties in the simulation are broadly consistent with observational constraints for the Milky Way \citep[e.g.][]{Bland-HawthornGerhard2016,HuntVasiliev2025}.

In Figure~\ref{fig:snapshot}(b), the mean radial velocity field $\langle v_R\rangle$ exhibits a pronounced quadrupole pattern aligned with the bar major axis at $R\lesssim 5~\mathrm{kpc}$.
At larger radii, the phase of the quadrupole progressively shifts in the trailing direction relative to the local rotation and forms a spiral-like pattern out to $R\sim 8~\mathrm{kpc}$.
Farther out, the velocity field departs from a simple quadrupole morphology, likely reflecting the influence of stellar spiral structure.
Similar non-axisymmetric streaming signatures have been reported in \textit{Gaia}-based analyses of the Milky Way disk.
These include bar-related quadrupole patterns in the inner Galaxy \citep[e.g.][]{Bovy+2019,Leung+2023} and spiral-arm--like features in maps of mean stellar velocity fields \citep[e.g.][]{Gaia-Katz+2018,Eilers+2020,Martinez-Medina+2022}.

Figure~\ref{fig:snapshot}(c) shows the residual mean azimuthal velocity field, $\langle v_\phi\rangle-\langle v_\phi\rangle_{\rm axi}$, where $\langle v_\phi\rangle_{\rm axi}$ is the azimuthally averaged mean azimuthal velocity at the same cylindrical radius.
The residual field exhibits a coherent bar-aligned pattern whose sign changes with radius.
In the bar region ($R\lesssim 5~\mathrm{kpc}$), azimuthal streaming is slower than the azimuthal average near the bar ends and faster along the bar sides.
In the inner disk outside the bar ($5\lesssim R\lesssim 8~\mathrm{kpc}$), the trend reverses, with faster streaming near the bar ends and slower streaming along the bar sides.

The detailed non-axisymmetric velocity pattern is not fixed in time and varies across snapshots, due to the time-dependent nature of the spiral structure itself and its interaction with the bar \citep[e.g.][]{SellwoodSparke1988,Baba2015c,Hilmi+2020}.
In the context of this paper, this time variability is treated as a source of snapshot-to-snapshot scatter in the inferred Jeans bias.
The implications for the systematic error budget are discussed in Section~\ref{subsec:results_terms}.

\subsection{Mock observational volume and reference axisymmetric circular speed}
\label{sec:method_mock}

We define a right-handed Cartesian coordinate system with the Galactic center at the origin and adopt cylindrical coordinates $(R,\phi,z)$.
The azimuth $\phi$ increases clockwise in the disk plane.
This is the direction of Galactic rotation when viewed from the north Galactic pole.
We place the mock observer (Sun) at $(x,y)=(0,-R_0)$ and adopt $R_0=8.2~\mathrm{kpc}$ \citep[][]{Bland-HawthornGerhard2016,GravityCollaboration+2021}.
We report radii in units of $R_0$.
Motivated by the Milky Way geometry, we adopt a Solar-like bar angle of $\phi_b=25^\circ$ \citep[][]{HuntVasiliev2025}.
Here $\phi_b$ is the angle of the bar major axis measured from the Sun--Galactic-center line in the same sense as $\phi$.

To mimic an observationally accessible region and to focus on stars close to the mid-plane, we select star particles in a wedge-shaped volume motivated by \citet{Eilers+2019}.
We require $|\phi|<30^\circ$, $R/R_0>0.5$, and $|z|<d\,\tan(6^\circ)$.
Here $d\equiv \sqrt{x^2+y^2}$ is the in-plane distance from the observer.
This mock observational volume and the observer location are indicated in Figure~\ref{fig:snapshot}.
All kinematic and density quantities used for the Jeans analysis (Section~\ref{sec:method_jeans}) are computed from this selected stellar sample.

For comparison with the Jeans-based estimate (Section~\ref{sec:method_jeans}), we define a reference axisymmetric circular speed $V_{\rm c,axi}(R)$ from the axisymmetric component of the gravitational potential.
We compute the gravitational potential using the \texttt{AGAMA} library \citep{Vasiliev2019}.
We model the total potential as the sum of components representing the dark-matter halo, a classical bulge, and the stellar and gas disks.
For the halo and classical bulge, we construct potential expansions using the spherical multipole representation (\texttt{Multipole}).
For the stellar and gas disks, we use the axisymmetric cylindrical-spline expansion (\texttt{CylSpline}) based on the corresponding particle distributions.
We then extract the axisymmetric (azimuthal Fourier $m=0$) component of the total potential, denoted $\Phi_0(R,z)$.
We define $V_{\rm c,axi}(R)$ from the radial gradient of $\Phi_0$ on the mid-plane ($z=0$) as
\begin{equation}
V_{\rm c,axi}^2(R) \equiv
R\,\left.\frac{\partial \Phi_0(R,z)}{\partial R}\right|_{z=0}.
\label{eq:vc_m0}
\end{equation}
Because the simulation does not necessarily reproduce the Milky Way circular-speed normalization at $R_0$, we normalize velocities by $V_0 \equiv V_{\rm c,axi}(R_0)$ when presenting rotation curves and residuals.

\subsection{Axisymmetric steady-state Jeans equation}
\label{sec:method_jeans}

\begin{figure*}
\begin{center}
\includegraphics[width=\linewidth]{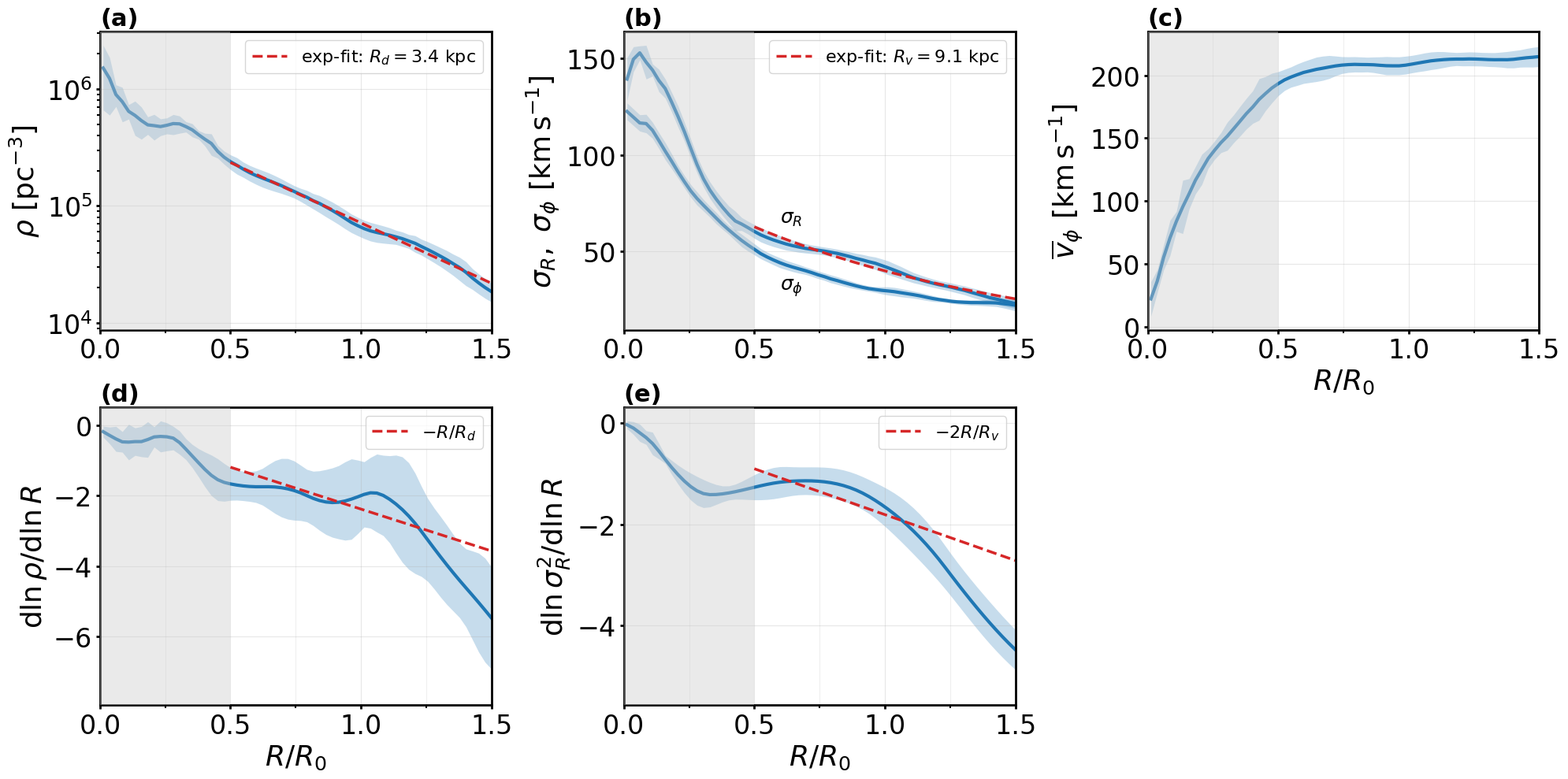}
\end{center}
\caption{
Radial profiles of the stellar kinematic and density quantities entering the axisymmetric Jeans analysis, measured in the mock observational volume and shown as functions of $R/R_0$.
\textbf{(a)} Number density of star particles, $\rho(R)$.
\textbf{(b)} Radial velocity dispersion, $\sigma_R(R)$, and azimuthal velocity dispersion, $\sigma_\phi(R)$.
\textbf{(c)} Mean azimuthal streaming velocity, $\overline{v}_\phi(R)$.
\textbf{(d)} Logarithmic density gradient, ${\rm d}\ln\rho/{\rm d}\ln R$.
\textbf{(e)} Logarithmic gradient of the squared radial dispersion, ${\rm d}\ln\sigma_R^2/{\rm d}\ln R$.
In each panel, the solid curve shows the snapshot-averaged profile.
The shaded band indicates the snapshot-to-snapshot scatter, shown as $\pm1\sigma$.
The dashed curves in panels (a) and (b) show exponential fits to $\rho(R)$ and $\sigma_R(R)$ over $0.5\le R/R_0 \le 1.5$, respectively.
The corresponding fitted gradients, $-R/R_{\rm d}$ and $-2R/R_{\rm v}$, are overplotted as dashed curves in panels (d) and (e).
The grey shaded region at $R/R_0<0.5$ marks radii excluded from the analysis
{\bf Alt text:}
Five-panel radial profiles of the tracer density and kinematic moments used in the axisymmetric Jeans analysis, showing the mean trends with radius and the snapshot-to-snapshot scatter; exponential-fit reference curves and the excluded inner region are indicated.
}
\label{fig:radial_profiles}
\end{figure*}

We begin with the Jeans equation for a collisionless stellar system in cylindrical coordinates $(R,\phi,z)$.
We assume a steady state ($\partial/\partial t = 0$) and axisymmetry ($\partial/\partial \phi = 0$).
Under these assumptions, the radial component reads
\begin{equation}
\rho\,\frac{\partial \Phi}{\partial R} =
\rho\,\frac{\overline{v_\phi^{\,2}}}{R}
-\frac{1}{R}\frac{\partial}{\partial R}\!\left(R\,\rho\,\overline{v_R^{\,2}}\right)
-\frac{\partial}{\partial z}\!\left(\rho\,\overline{v_R v_z}\right),
\label{eq:jeans_R_cyl_axisym}
\end{equation}
where $\rho$ is the tracer number density and $\Phi$ is the gravitational potential.
The overbar denotes an average over the local velocity distribution \citep[][]{BinneyTremaine2008}.
Here $(v_R,v_\phi,v_z)$ are the cylindrical velocity components.

To infer the circular speed from stellar kinematics, we adopt two additional approximations.
First, we assume negligible mean meridional streaming, so that $\overline{v_R}\simeq 0$ and $\overline{v_z}\simeq 0$.
Second, we neglect the tilt term near the mid-plane, so that
$\partial(\rho\,\overline{v_R v_z})/\partial z \approx 0$ at $z\simeq 0$.
These assumptions are motivated by our tracer selection, which is restricted to a thin region around the mid-plane ($|z/d|\le\tan 6^\circ$; Section~\ref{sec:method_sim}).
They are also commonly adopted in observational Jeans analyses of the Milky Way disk \citep[e.g.][]{Eilers+2019,Poder+2023}.
In the simulation, bar-driven streaming produces regions with non-zero mean radial motions in the gridded map (Figure~\ref{fig:snapshot}b).
Our Jeans estimator follows common observational assumptions and sets $\overline{v_R}\simeq 0$ within the mock wedge.
We therefore include any impact of non-zero radial streaming in the total bias measured in Section~\ref{sec:Results}, rather than trying to quantify the contribution from $\overline{v_R}$ separately.

We define the radial and azimuthal velocity dispersions as the second central moments,
$\sigma_R^2(R)\equiv \overline{v_R^{\,2}}-\overline{v_R}^{\,2}$ and
$\sigma_\phi^2(R)\equiv \overline{v_\phi^{\,2}}-\overline{v_\phi}^{\,2}$.
We define the circular speed on the mid-plane as $V_{\rm c}^2(R)\equiv R\,\partial\Phi(R,0)/\partial R$.
With these definitions, equation~(\ref{eq:jeans_R_cyl_axisym}) reduces to the standard axisymmetric, steady-state Jeans estimator
\begin{equation}
V_{\rm c,Jeans}^2(R) \approx
\underbrace{\overline{v_\phi}^{\,2}(R)}_{\rm streaming}
+
\underbrace{\sigma_\phi^2(R)}_{\rm azimuthal~dispersion}
+
\underbrace{T_{\rm AD}(R)}_{\rm asymmetric~drift},
\label{eq:jeans_axisym}
\end{equation}
where the three contributions are the azimuthal streaming term $\overline{v_\phi}^{\,2}$, the azimuthal-dispersion term $\sigma_\phi^{2}$, and the asymmetric-drift correction term $T_{\rm AD}$. 
The asymmetric-drift term provides the pressure-support correction from random motions, so that $\overline{v_\phi}$ can be lower than the circular speed even in equilibrium \citep[][]{BinneyTremaine2008}.
The asymmetric-drift term is
\begin{equation}
T_{\rm AD}(R)\equiv
-\sigma_R^2(R)\left(
1+\frac{\partial\ln\rho}{\partial\ln R}
+\frac{\partial\ln\sigma_R^2}{\partial\ln R}
\right).
\label{eq:Tad_def}
\end{equation}
In practice, we measure all moments within our thin wedge around the mid-plane (Section~\ref{sec:method_sim}).
We treat these measurements as representative of $z\simeq 0$.

We measure the radial profiles of $\rho(R)$, $\overline{v_\phi}(R)$, $\sigma_R(R)$, and $\sigma_\phi(R)$ from the selected stellar sample.
We use the same radial binning for all moments.
Figure~\ref{fig:radial_profiles} summarizes the snapshot-averaged profiles and the snapshot-to-snapshot scatter.
The density and radial velocity dispersion decline with radius in an approximately exponential manner.
In contrast, $\overline{v_\phi}(R)$ rises gradually in the inner disk and is nearly flat at larger radii.
At $R\simeq R_0$, we find $\sigma_R \approx 40~\mathrm{km\,s^{-1}}$, which is comparable to observational constraints for Milky Way disk stars \citep[e.g.][]{Bland-HawthornGerhard2016,Sharma+2021a}.

A key practical issue in applying equation~(\ref{eq:jeans_axisym}) is the evaluation of $T_{\rm AD}(R)$.
The main sensitivity arises from the logarithmic gradients $\partial \ln \rho/\partial \ln R$ and $\partial \ln \sigma_R^2/\partial \ln R$.
We consider two approaches.
First, we follow common observational practice and fit exponential models to the binned profiles over $0.5\le R/R_0\le 1.5$ \citep[e.g.][]{Eilers+2019}.
We adopt $\rho(R)\propto \exp(-R/R_{\rm d})$ and $\sigma_R(R)\propto \exp(-R/R_{\rm v})$.
Here $R_{\rm d}$ and $R_{\rm v}$ are the scale lengths of the tracer density and the radial velocity dispersion.
With these fits, the gradients become ${\rm d}\ln\rho/{\rm d}\ln R=-R/R_{\rm d}$ and ${\rm d}\ln\sigma_R^2/{\rm d}\ln R=-2R/R_{\rm v}$.
Substituting them into equation~(\ref{eq:Tad_def}) gives
\begin{equation}
T_{\rm AD}^{\rm (fit)}(R)=
-\sigma_R^2(R)\left(
1-\frac{R}{R_{\rm d}}-2\,\frac{R}{R_{\rm v}}
\right).
\label{eq:Tad_exp}
\end{equation}
Combining equation~(\ref{eq:Tad_exp}) with equation~(\ref{eq:jeans_axisym}) yields the exponential-fit Jeans estimate, $V_{\rm c,Jeans}^{\rm (fit)}(R)$.
Figure~\ref{fig:radial_profiles}(d,e) shows the corresponding gradients as dashed curves.

Second, we evaluate the gradients numerically from the binned profiles to assess sensitivity to the derivative treatment.
Direct finite differencing can be noisy.
We therefore smooth $\rho(R)$ and $\sigma_R(R)$ using a smoothing spline implemented by \texttt{UnivariateSpline} in \texttt{scipy}.
We adopt a smoothing parameter $s=N_{\rm data}$, where $N_{\rm data}$ is the number of radial bins.
We then differentiate the smoothed profiles to obtain $\partial \ln \rho/\partial \ln R$ and $\partial \ln \sigma_R^2/\partial \ln R$.
We compute $T_{\rm AD}^{\rm (num)}(R)$ by inserting these gradients into equation~(\ref{eq:Tad_def}), and obtain a second Jeans estimate, $V_{\rm c,Jeans}^{\rm (num)}(R)$. Figure~\ref{fig:radial_profiles}(d,e) shows the numerically evaluated gradients.

\subsection{Snapshot ensemble and empirical bias distribution}
\label{subsec:method_biasdist}

For each snapshot in the time window $t=2$--$3~\mathrm{Gyr}$, we repeat the full measurement procedure and compare the Jeans-based estimates with the reference axisymmetric curve. We evaluate the asymmetric-drift term using two derivative treatments. One uses exponential fits to the binned profiles (fit), and the other uses numerically evaluated derivatives (num).

For each treatment, we quantify the bias by the fractional residual
\begin{equation}
\delta V_{\rm c}(R)\equiv
\frac{V_{\rm c,Jeans}(R)-V_{\rm c,axi}(R)}{V_{\rm c,axi}(R)},
\label{eq:deltaVc_def}
\end{equation}
where $V_{\rm c,Jeans}(R)$ denotes either $V_{\rm c,Jeans}^{\rm (fit)}(R)$ or $V_{\rm c,Jeans}^{\rm (num)}(R)$.
Within each radial bin, we combine $\delta V_{\rm c}$ values from all snapshots to construct an empirical bias distribution at fixed $R$.
We repeat this procedure for both treatments.
We summarize each distribution by its mean $\mu_{\rm sys}(R)$ and dispersion $\sigma_{\rm sys}(R)$.
These quantities represent the systematic bias and the snapshot-to-snapshot scatter, respectively.

We emphasize that any assessment based on a single Milky Way analogue is model dependent.
If the simulation differs systematically from the real Milky Way, additional systematics may arise that are not captured by the present experiment.
Examples include differences in the bar or spiral amplitudes, the pattern speed, and the equilibrium velocity-dispersion structure.
We therefore interpret our results as a controlled estimate of the bias induced by non-axisymmetric streaming motions in a self-consistent barred-disk simulation.
We do not present it as a complete error budget for the real Milky Way.

\section{Results}
\label{sec:Results}

\subsection{Bias and scatter in Jeans-based circular-speed estimates}
\label{subsec:results_bias}

We quantify the bias and snapshot-to-snapshot scatter that arise when the steady-state, axisymmetric Jeans equation is applied to the mock observational volume described in Section~\ref{sec:methods}.
For each snapshot in our post-bar-formation sample (Section~\ref{sec:method_sim}), we compute $V_{\rm c,Jeans}(R)$ and compare it with the reference axisymmetric circular-speed curve $V_{\rm c,axi}(R)$ defined in equation~(\ref{eq:vc_m0}).

\begin{figure*}
\begin{center}
\includegraphics[width=\linewidth]{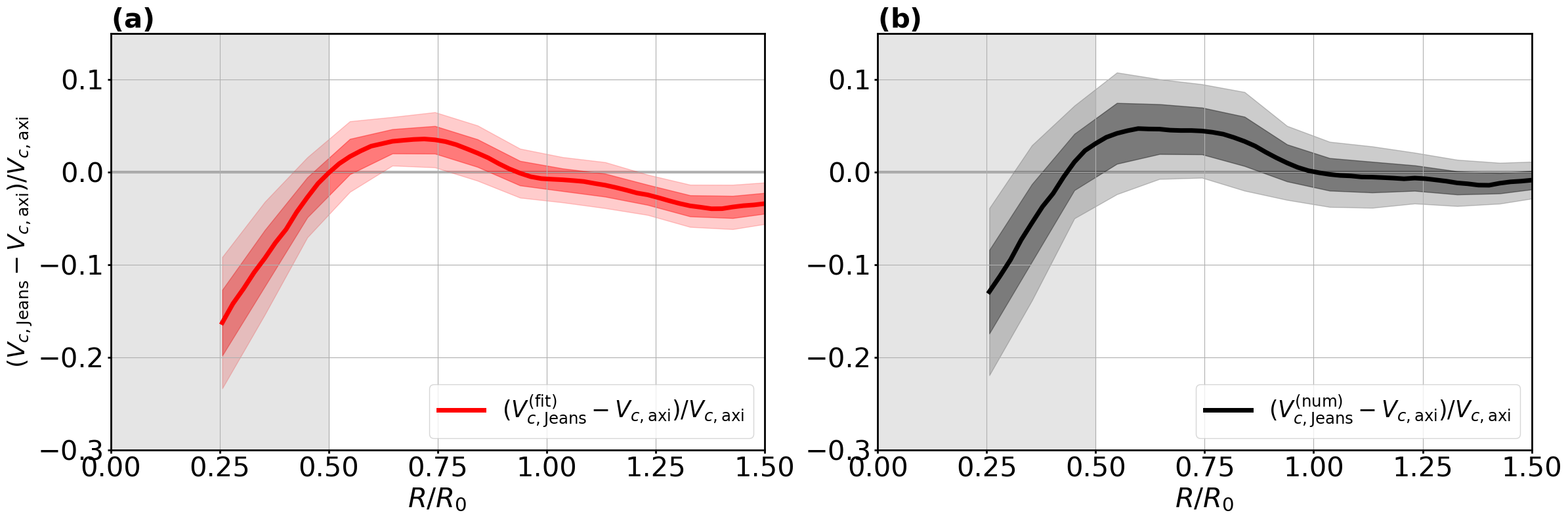}
\end{center}
\caption{
Fractional residual of the Jeans-inferred circular speed relative to the reference axisymmetric circular speed,
$\delta V_{\rm c}(R)\equiv \bigl[V_{\rm c,Jeans}(R)-V_{{\rm c},{\rm axi}}(R)\bigr]/V_{{\rm c},{\rm axi}}(R)$, shown as a function of radius.
The horizontal axis shows radius normalized by $R_0$.
The thick curve gives the snapshot-averaged residual, while the shaded bands show the $1\sigma_{\rm sys}$ and $2\sigma_{\rm sys}$ ranges of the snapshot-to-snapshot distribution at each radius.
\textbf{(a)}: exponential-fit treatment of the logarithmic gradients in the Jeans equation.
\textbf{(b)}: numerical-derivative treatment, where the gradients are obtained by differentiating smoothed radial profiles.
The gray shaded region at $R/R_0<0.5$ marks the radial range excluded from the analysis.
{\bf Alt text:}
Two-panel plot of the fractional difference between the Jeans-inferred and axisymmetric reference circular speeds versus normalized radius, showing the snapshot-averaged bias and the 1$\sigma$ and 2$\sigma$ scatter bands for exponential-fit (left) and numerical-derivative (right) treatments.
}
\label{fig:rotation_carve}
\end{figure*}

Figure~\ref{fig:rotation_carve} shows the fractional residual $\delta V_{\rm c}(R)$ defined in equation~(\ref{eq:deltaVc_def}).
The thick horizontal line marks $\delta V_{\rm c}(R)=0$, which corresponds to perfect agreement with $V_{\rm c,axi}(R)$.
The thick red curve shows the snapshot-averaged $\delta V_{\rm c}(R)$.
At each radius, we summarize the snapshot-to-snapshot distribution by its mean $\mu_{\rm sys}(R)$ and dispersion (standard deviation) $\sigma_{\rm sys}(R)$, obtained from a Gaussian fit (described below).
The shaded bands show the $\pm1\sigma_{\rm sys}$ and $\pm2\sigma_{\rm sys}$ ranges around $\mu_{\rm sys}$.
Panel~(a) uses the exponential-fit treatment $V_{\rm c,Jeans}^{\rm (fit)}$, and panel~(b) uses the numerical-derivative treatment $V_{\rm c,Jeans}^{\rm (num)}$.
The two treatments differ only in how the logarithmic gradients entering the asymmetric-drift term $T_{\rm AD}$ are evaluated.
All other Jeans moments are measured from the same tracer sample in the same way.

In the inner disk ($0.5\lesssim R/R_0 \lesssim 1.0$), the mean residual is systematically positive for both treatments.
This indicates that the axisymmetric Jeans model overestimates $V_{\rm c,axi}$ in this region.
Across most of this range, the numerical-derivative treatment yields a larger mean residual than the exponential-fit treatment.
The overall radial trend is similar between the two treatments.
This shows that the presence of an inner-disk overestimate is robust, while its amplitude depends on the treatment of the gradients in $T_{\rm AD}$.

\begin{figure*}
\begin{center}
\includegraphics[width=\linewidth]{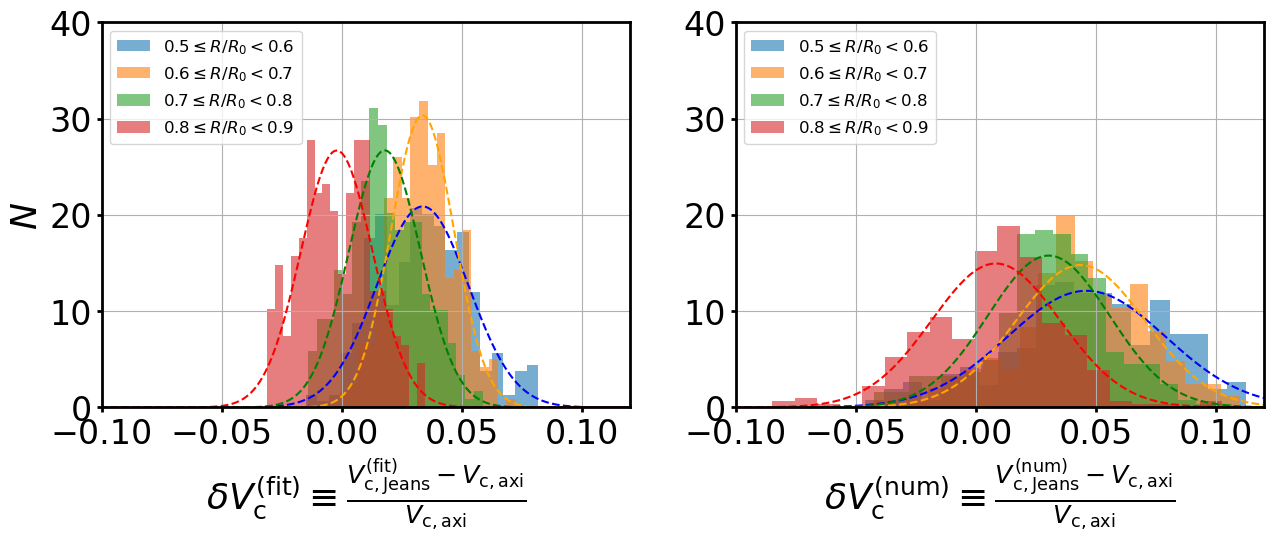}
\end{center}
\caption{
Histograms of the fractional residual $\delta V_{\rm c}(R)\equiv \bigl[V_{\rm c,Jeans}(R)-V_{{\rm c},{\rm axi}}(R)\bigr]/V_{{\rm c},{\rm axi}}(R)$ in several radial bins, constructed by pooling measurements from all snapshots.
In each bin, the dashed curve shows the best-fitting Gaussian distribution, from which we extract the mean bias $\mu_{\rm sys}(R)$ and scatter $\sigma_{\rm sys}(R)$ used throughout the analysis.
\textbf{(Left)}: exponential-fit treatment of the logarithmic gradients in the Jeans equation.
\textbf{(Right)}: numerical-derivative treatment, where the gradients are obtained by differentiating smoothed radial profiles.
{\bf Alt text:}
Two-panel set of histograms of the fractional circular-speed residuals in multiple radial bins (pooled over snapshots), with best-fitting Gaussian curves overplotted; left panel uses exponential-fit gradients and right panel uses numerical-derivative gradients, illustrating the mean bias and scatter in each bin.
}
\label{fig:err_hist}
\end{figure*}

To characterize the inner-disk bias statistically, we examine the empirical distribution of $\delta V_{\rm c}(R)$ in radial bins.
We adopt a bin width of $\Delta(R/R_0)=0.1$.
Within each bin, we compile $\delta V_{\rm c}$ values from all snapshots.
We then perform Gaussian fits to the resulting distributions and estimate $\mu_{\rm sys}$ and $\sigma_{\rm sys}$ for both treatments.
Figure~\ref{fig:err_hist} shows the resulting histograms for $0.5\le R/R_0 \le 0.9$.
The bias is strongest around $R/R_0\simeq 0.6$.
In this bin, the exponential-fit treatment gives $\mu_{\rm sys}^{\rm (fit)}\simeq 0.034$ with $\sigma_{\rm sys}^{\rm (fit)}\simeq 0.019$.
The numerical-derivative treatment yields a larger mean bias, with $\mu_{\rm sys}^{\rm (num)}\simeq 0.046$ and $\sigma_{\rm sys}^{\rm (num)}\simeq 0.033$.
This larger scatter likely reflects the fact that the numerical-derivative treatment tracks time-dependent changes in ${\rm d}\ln\rho/{\rm d}\ln R$ and ${\rm d}\ln\sigma_R^2/{\rm d}\ln R$ (see Figure~\ref{fig:radial_profiles}d and e).
These gradients enter $T_{\rm AD}$ and therefore amplify snapshot-to-snapshot variability in $V_{\rm c,Jeans}$.
Toward larger radii, the mean bias decreases and is consistent with zero within $\lesssim 1\%$ by $R/R_0\simeq 0.9$ for both treatments.
Overall, the mean bias is positive and corresponds to an offset of $\sim 1.5$--$2\,\sigma_{\rm sys}$ relative to the snapshot-to-snapshot scatter, depending on the derivative treatment.

As a secondary point, Figure~\ref{fig:rotation_carve} shows modest differences between the two treatments at larger radii.
With the exponential-fit treatment, the mean residual becomes slightly negative.
With the numerical-derivative treatment, the mean residual remains consistent with $\delta V_{\rm c}\simeq 0$ within the snapshot-to-snapshot scatter.
This behaviour is consistent with the tracer density and dispersion profiles deviating from a single exponential over the fitted range (Figure~\ref{fig:radial_profiles}a,b).
In our mock volume, the outer-disk profiles are shallower than the inner-disk trend.
A single-exponential fit can therefore bias the logarithmic gradients entering $T_{\rm AD}$ and shift $V_{\rm c,Jeans}^{\rm (fit)}$.

The apparent outer-disk agreement of the numerical-derivative treatment should not be over-interpreted.
Our isolated-galaxy simulation does not include external perturbations such as those from the Sagittarius dwarf galaxy.
In the real Milky Way, such perturbations can be important at large radii \citep[e.g.][]{Laporte+2019,Hunt+2021,Asano+2025}.
They may also produce systematic offsets between Jeans-based estimates and the underlying axisymmetric circular speed \citep[][]{Koop+2024}.

\subsection{Term budget of the Jeans estimator}
\label{subsec:results_terms}

\begin{figure}
\begin{center}
\includegraphics[width=\linewidth]{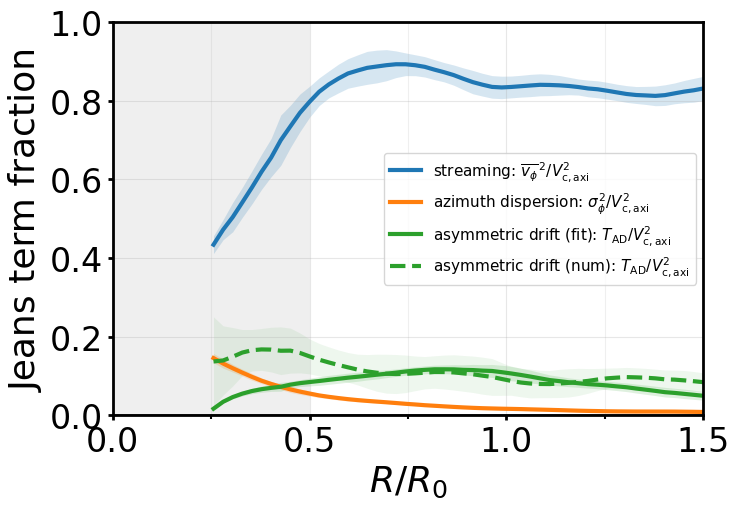}
\end{center}
\caption{
Radial profiles of the relative contributions of the individual terms in the axisymmetric Jeans estimator (equation~\ref{eq:jeans_axisym}), \textbf{normalized by the squared reference axisymmetric circular speed $V_{\rm c,axi}^2$}.
The streaming term $\overline{v_\phi}^{\,2}$ (blue), the azimuthal-dispersion term $\sigma_\phi^{2}$ (orange), and the asymmetric-drift term $T_{\rm AD}$ (green) are shown.
Solid curves use the exponential-fit evaluation of the logarithmic gradients in $T_{\rm AD}$, while dashed curves use the numerical-derivative evaluation.
Shaded bands indicate the propagated $1\sigma$ uncertainties.
The uncertainty band for the azimuthal-dispersion term is also shown, but it is difficult to see because the $\sigma_\phi^{2}$ fraction is small over the plotted range.
The gray shaded region at $R/R_0<0.5$ marks the radial range excluded from the analysis.
{\bf Alt text:}
Single-panel radial plot of the fractional contributions to \textbf{$V_{\rm c,axi}^2$} from the streaming term $\overline{v_\phi}^{\,2}$, the azimuthal-dispersion term $\sigma_\phi^{2}$, and the asymmetric-drift term $T_{\rm AD}$, shown as solid (exponential-fit) and dashed (numerical-derivative) curves with shaded $1\sigma$ uncertainty bands; radii $R/R_0<0.5$ are shaded as excluded.
}
\label{fig:jeans_terms}
\end{figure}

To clarify which ingredients drive the inner-disk bias, we decompose the axisymmetric Jeans estimator (equation~\ref{eq:jeans_axisym}) into its individual terms. At each radius, we express the contribution of each term as a fraction of the squared reference axisymmetric circular speed, i.e.\ the term divided by $V_{\rm c,axi}^2$.
Figure~\ref{fig:jeans_terms} shows the resulting radial profiles.
Over $0.5\lesssim R/R_0\lesssim 1.5$, the Jeans estimator is dominated by the azimuthal streaming contribution $\overline{v_\phi}^{\,2}$.
The azimuthal-dispersion contribution $\sigma_\phi^{2}$ is small over the same range and remains sub-dominant.
The asymmetric-drift term $T_{\rm AD}$ contributes at the $\sim 10\%$ level over $0.5\lesssim R/R_0\lesssim 1.5$.
Since $T_{\rm AD}$ enters equation~(\ref{eq:jeans_axisym}) with a positive sign, it increases $V_{\rm c,Jeans}^{2}$.
The solid and dashed curves differ only in how the logarithmic gradients entering $T_{\rm AD}$ are evaluated, so the treatment dependence is seen most clearly in $T_{\rm AD}$ itself.
In the inner disk, the difference between the two $T_{\rm AD}$ evaluations is modest, consistent with the broadly similar inner-disk residual trends in Figure~\ref{fig:rotation_carve}.

\begin{figure*}
\begin{center}
\includegraphics[width=\linewidth]{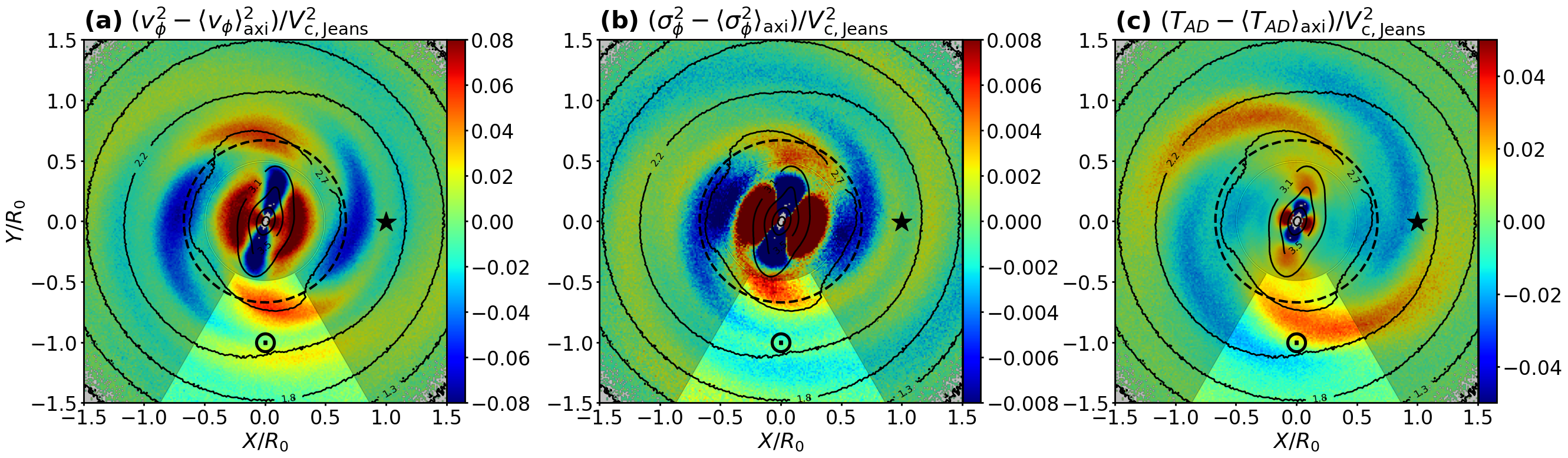}
\end{center}
\caption{
Snapshot-averaged residual maps of the Jeans ingredients relative to their axisymmetric reference profiles.
Panels~(a)--(c) show the residuals of the streaming term $\overline{v_\phi}^{\,2}$, the azimuthal-dispersion term $\sigma_\phi^{2}$, and the asymmetric-drift term $T_{\rm AD}$, respectively.
At each position, the local value of each term is compared with its azimuthally averaged radial profile at the same cylindrical radius $R$.
The residuals are normalized by the squared Jeans-based circular speed $V_{\rm c,Jeans}^2(R)$ at the same radius. The corotation radius ($R_{\rm CR}/R_0\approx 0.67$) is overplotted as a dashed circle.
The observer position used for the Figure~\ref{fig:rotation_carve_90} analysis is marked by a star symbol.
{\bf Alt text:}
Three face-on $(x,y)$ maps showing snapshot-averaged, normalized residuals relative to the azimuthally averaged radial profiles for (a) $\overline{v_\phi}^{\,2}$, (b) $\sigma_\phi^{2}$, and (c) $T_{\rm AD}$; each residual is scaled by $V_{\rm c,Jeans}^2(R)$ at the same cylindrical radius, and a dashed circle marks the corotation radius at $R_{\rm CR}/R_0\simeq 0.67$.
}
\label{fig:resi_map}
\end{figure*}

Figure~\ref{fig:resi_map} then visualizes the corresponding azimuthal structure across the disk plane.
The residual fields are stacked over the full snapshot ensemble.
The shaded regions mark the area outside the mock observational wedge used for the Jeans measurements, defined in the same way as in Figure~\ref{fig:snapshot}.
Panels~(a)--(c) show the residuals of the streaming term $\overline{v_\phi}^{\,2}$, the azimuthal-dispersion term $\sigma_\phi^{2}$, and the asymmetric-drift term $T_{\rm AD}$ relative to their axisymmetric reference profiles at the same cylindrical radius.
In each panel, the residual is normalized by $V_{\rm c,Jeans}^2$ at the corresponding radius.
Within the wedge, all three ingredients exhibit coherent azimuthal structure in the inner disk.

Combined with the term budget in Figure~\ref{fig:jeans_terms}, Figure~\ref{fig:resi_map} shows that the mean bias is set primarily by coherent azimuthal variations in the streaming field.
In the snapshot-averaged maps, the normalized residual amplitude is largest for $\overline{v_\phi}^{\,2}$ (typically $\sim 0.05$ in units of $V_{\rm c,Jeans}^2$), intermediate for $T_{\rm AD}$ ($\sim 0.03$), and much smaller for $\sigma_\phi^{2}$ ($\sim 0.005$).
Accordingly, the dominant modulation is carried by the streaming term, while $T_{\rm AD}$ provides a non-negligible secondary contribution.
In the Solar-like configuration ($\phi_b=25^\circ$) over $0.5\lesssim R/R_0\lesssim 1$, the wedge preferentially samples locations where the residuals in $\overline{v_\phi}^{\,2}$ and $T_{\rm AD}$ share the same sign and therefore add constructively.
This picture is further supported by the morphology of the streaming residual itself: the snapshot-averaged map of $\overline{v_\phi}^{\,2}$ displays a clean, bar-aligned $m=2$ (quadrupole) pattern.
The phase of this quadrupole-like residual shifts by $90^\circ$ across the corotation radius ($R/R_0\approx 0.67$; dashed circle), as expected for bar-driven responses on opposite sides of corotation \citep[][]{BinneyTremaine2008}.
A closely similar quadrupole pattern in residual azimuthal streaming is seen in test-particle simulations in an imposed bar potential \citep[e.g.][]{Monari+2016b}.

\begin{figure*}
\begin{center}
\includegraphics[width=\linewidth]{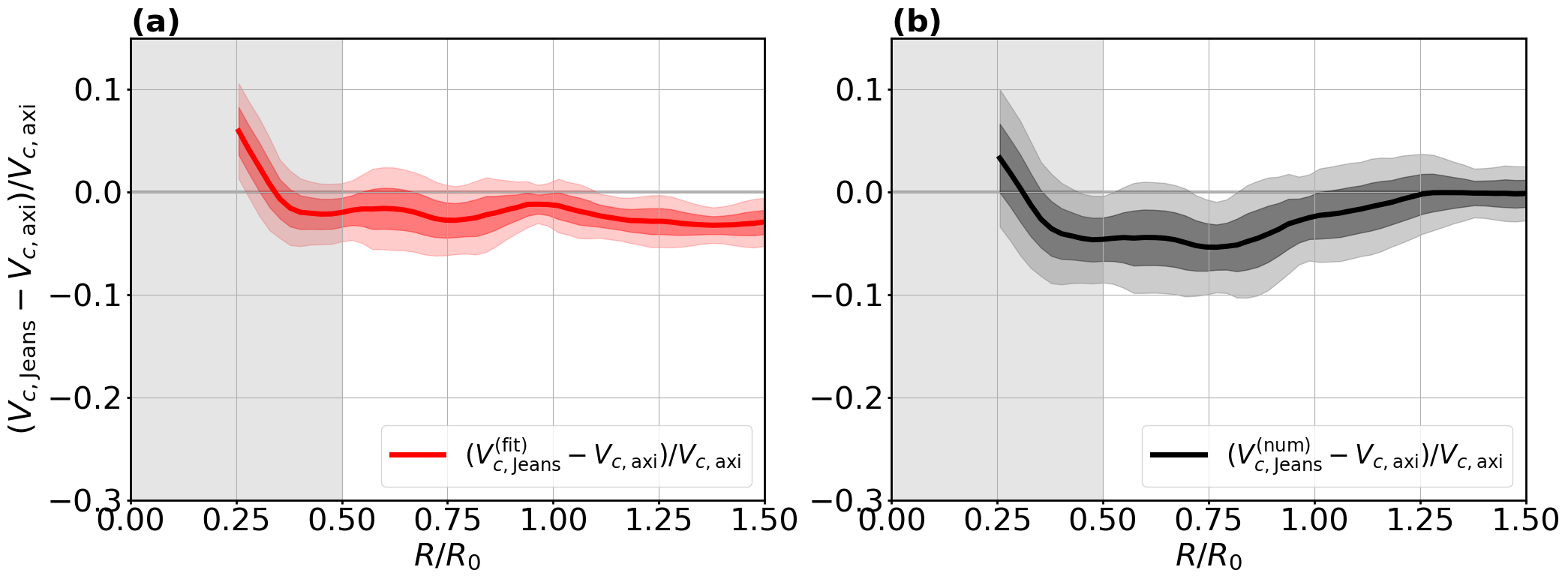}
\end{center}
\caption{
Same as Figure~\ref{fig:rotation_carve}, but for the observer configuration with $\phi_{\rm b}=115^\circ$, where $\phi_{\rm b}$ is the observer azimuth measured from the bar major axis.
The $\phi_{\rm b}=115^\circ$ observer position is marked by a star symbol in Figure~\ref{fig:resi_map} to guide the interpretation of the viewing geometry.
{\bf Alt text:}
Two-panel plot of the fractional residual $\delta V_{\rm c}(R)$ versus $R/R_0$ for the $\phi_{\rm b}=115^\circ$ observer configuration (azimuth measured from the bar major axis), showing the snapshot-averaged curve with surrounding $1\sigma_{\rm sys}$ and $2\sigma_{\rm sys}$ bands for the exponential-fit (left) and numerical-derivative (right) treatments.
}
\label{fig:rotation_carve_90}
\end{figure*}

Figure~\ref{fig:rotation_carve_90} shows the corresponding result for the  $\phi_b=115^\circ$ configuration.
In this viewing geometry, the mean residual in the inner disk reverses sign relative to the Solar-like case, and $V_{\rm c,Jeans}$ tends to underestimate $V_{\rm c,axi}$.
This behaviour follows naturally from the same bar-driven $m=2$ streaming pattern: rotating the observer azimuth from the Solar-like $\phi_b=25^\circ$ to $\phi_b=115^\circ$ moves the wedge from the ``fast-streaming'' sector to the ``slow-streaming'' sector of the quadrupole residual.
As a result, the dominant streaming contribution shifts in the opposite sense and the net Jeans bias changes sign.

Finally, Figure~\ref{fig:resi_map} characterizes the snapshot-averaged residual pattern, while the remaining snapshot-to-snapshot scatter around this mean bias reflects time variability in the non-axisymmetric velocity field.
The scatter is largest in the radial range where the disk shows prominent spiral-like streaming patterns (Figure~\ref{fig:snapshot}c), suggesting that transient spiral structure contributes to the time-dependent component of the Jeans bias.
In barred disks, such variability is commonly associated with transient, recurrent spirals and their interaction with the bar \citep[e.g.][]{SellwoodSparke1988,Grand+2012b,Baba2015c,Hilmi+2020,Vislosky+2024}.

\section{Discussion}
\label{sec:Discussion}

This paper used a controlled simulation-based experiment to quantify the systematic offset that can arise when a steady, axisymmetric Jeans analysis is applied to a barred disk with coherent non-circular motions.
In a Solar-like configuration, the Jeans-inferred circular-speed curve is biased high in the inner disk by about $4\%$ (corresponding to $\approx 10~\mathrm{km\,s^{-1}}$).
The mean offset is significant relative to the snapshot-to-snapshot scatter.
The bias can reverse sign for a different observer azimuth, which shows that the inferred rotation curve depends on viewing geometry in a barred disk.

A key implication is that different Jeans-based rotation curves can agree closely with each other while still being systematically offset from the axisymmetrically defined circular-speed curve, because they may share the same modelling assumptions and survey geometry.
In the inner disk, recent Jeans analyses based on \textit{Gaia} data show very good mutual consistency, with differences at the level of only a few $\mathrm{km\,s^{-1}}$ \citep[see Fig.~17 of][]{HuntVasiliev2025}.
In light of the simulation-based experiment presented here, those consistent Jeans rotation curves could still share a common absolute bias.
In particular, they may overestimate the axisymmetric circular speed by $\approx 4\%$ (i.e. $\approx 10~\mathrm{km\,s^{-1}}$), if the surveyed wedge samples a streaming field that is coherently offset from the azimuthal average.
As a scaling under the usual spherical approximation, a $\approx 4\%$ bias in circular speed corresponds to a $\approx 8\%$ overestimate of the enclosed dynamical mass at fixed radius.

The term budget and residual maps support a simple physical interpretation.
In the inner disk, $V_{\rm c,Jeans}^2$ is dominated by the azimuthal streaming contribution, while the other terms contribute at a lower level.
As a result, a coherent offset in $\overline{v_\phi}$ within a restricted azimuthal selection propagates directly into the Jeans-based circular-speed estimate.
In other words, the leading limitation is not random measurement uncertainty, but a geometry-driven systematic caused by coherent non-axisymmetric streaming in combination with limited sky coverage.

A practical point is that the inner-disk offset is not removed by changing the treatment of the asymmetric-drift gradients.
The exponential-fit and numerical-derivative treatments differ only in how the logarithmic gradients entering $T_{\rm AD}$ are evaluated.
Both treatments yield a positive mean residual over $0.5\lesssim R/R_0\lesssim 1.0$, although the numerical-derivative treatment gives a larger mean bias.
It also yields a larger scatter because it tracks time-dependent changes in ${\rm d}\ln\rho/{\rm d}\ln R$ and ${\rm d}\ln\sigma_R^2/{\rm d}\ln R$.
These gradients enter $T_{\rm AD}$ and therefore amplify snapshot-to-snapshot variability in $V_{\rm c,Jeans}(R)$.

Although the discussion above focuses on $0.5\lesssim R/R_0\lesssim 1.5$, Figure~\ref{fig:rotation_carve} (and Figure~\ref{fig:rotation_carve_90}) also shows that the axisymmetric Jeans estimator can be far more strongly biased inside the bar region ($R\lesssim 5~\mathrm{kpc}$).
At $R/R_0\approx 0.25$, the $\phi_{\rm b}=25^\circ$ (Solar-like) configuration yields a large negative residual, with $V_{\rm c,Jeans}$ underestimating $V_{\rm c,axi}$ by $\approx 10$--$20\%$, whereas the $\phi_{\rm b}=115^\circ$ configuration shows the opposite trend, overestimating $V_{\rm c,axi}$ by $\approx 5$--$10\%$.
While precise 6D stellar phase-space information in the bar region is observationally challenging, these results suggest that even if such data become available in the future, applying a steady, axisymmetric Jeans analysis without modelling non-axisymmetric streaming could introduce order-of-ten-percent biases in the inferred circular-speed curve in the bar-dominated inner Milky Way.

This analysis is based on a single Milky Way analogue, so the numerical values are model dependent.
In particular, in $N$-body disk simulations the amplitude and persistence of spiral structure can depend on finite-$N$ noise \citep[][]{Fujii+2011,Fujii+2019}. Our simulation uses $\sim 1.2\times 10^{7}$ star particles, placing it in a high-resolution regime discussed in these studies, although the detailed spiral strength may still vary with resolution.
Nevertheless, the key inner-disk result is that, for a Solar-like viewing geometry, bar-driven streaming produces a coherent offset in $\overline{v_\phi}$ within the observed wedge, which sets the mean Jeans bias, while time-dependent spirals mainly contribute to the snapshot-to-snapshot dispersion around that mean.
To account for this geometry-driven bias, Jeans-based rotation-curve analyses in the inner disk should either incorporate an explicit non-axisymmetric streaming model (e.g.\ by fitting for bar-driven azimuthal structure), or calibrate and marginalize over a systematic offset using suites of barred Milky Way analogues spanning plausible bar/spiral parameters and viewing geometries.
This would turn the geometric bias identified here into a quantified uncertainty that can be propagated into bulge--bar--disk decompositions and inner-disk mass constraints.

\section*{Funding}
JB was supported by the Japan Society for the Promotion of Science (JSPS) under Grant Numbers 24K07095, 25H00394, and 25K24687.

\section*{Data availability} 
The simulation snapshots are available upon request. 

\begin{ack}
We sincerely thank the anonymous referee for their constructive comments, which helped improve the clarity and context of this paper.
We thank Kohei Hattori, Keiichi Wada and Atsushi Tanimoto for helpful discussions and comments.
Calculations, numerical analyses, and visualization were carried out on Cray XD2000 (ATERUI-III) and computers at the Center for Computational Astrophysics, National Astronomical Observatory of Japan (CfCA/NAOJ). 
\end{ack}


\begin{thebibliography}{63}
\expandafter\ifx\csname natexlab\endcsname\relax\def\natexlab#1{#1}\fi

\bibitem[{{Almannaei} {et~al}\mbox{.}(2024){Almannaei}, {Kawata}, {Baba}, {Hunt}, {Seabroke}, \& {Yan}}]{Almannaei+2024}
{Almannaei} A.~S., {Kawata} D., {Baba} J., {Hunt} J. A.~S., {Seabroke} G., {Yan} Z., 2024, \mnras, 529, 1035

\bibitem[{{Antoja} {et~al}\mbox{.}(2018){Antoja}, {Helmi}, {Romero-G{\'o}mez}, {Katz}, {Babusiaux}, {Drimmel}, {Evans}, {Figueras}, {Poggio}, {Reyl{\'e}}, {Robin}, {Seabroke}, \& {Soubiran}}]{Antoja+2018}
{Antoja} T. {et~al.}, 2018, \nat, 561, 360

\bibitem[{{Asano} {et~al}\mbox{.}(2020){Asano}, {Fujii}, {Baba}, {B{\'e}dorf}, {Sellentin}, \& {Portegies Zwart}}]{Asano+2020}
{Asano} T., {Fujii} M.~S., {Baba} J., {B{\'e}dorf} J., {Sellentin} E., {Portegies Zwart} S., 2020, \mnras, 499, 2416

\bibitem[{{Asano} {et~al}\mbox{.}(2025){Asano}, {Fujii}, {Baba}, {Portegies Zwart}, \& {B{\'e}dorf}}]{Asano+2025}
{Asano} T., {Fujii} M.~S., {Baba} J., {Portegies Zwart} S., {B{\'e}dorf} J., 2025, \aap, 700, A109

\bibitem[{{Baba}(2015)}]{Baba2015c}
{Baba} J., 2015, \mnras, 454, 2954

\bibitem[{{Baba}(2025{\natexlab{a}})}]{Baba2025b}
{Baba} J., 2025{\natexlab{a}}, \apj, 989, 121

\bibitem[{{Baba}(2025{\natexlab{b}})}]{Baba2025a}
{Baba} J., 2025{\natexlab{b}}, \pasj, 77, 916

\bibitem[{{Baba} {et~al}\mbox{.}(2018){Baba}, {Kawata}, {Matsunaga}, {Grand}, \& {Hunt}}]{Baba+2018}
{Baba} J., {Kawata} D., {Matsunaga} N., {Grand} R. J.~J., {Hunt} J. A.~S., 2018, \apjl, 853, L23

\bibitem[{{Binney} \& {Tremaine}(2008)}]{BinneyTremaine2008}
{Binney} J., {Tremaine} S., 2008, {Galactic Dynamics: Second Edition}. Princeton University Press

\bibitem[{{Bland-Hawthorn} \& {Gerhard}(2016)}]{Bland-HawthornGerhard2016}
{Bland-Hawthorn} J., {Gerhard} O., 2016, \araa, 54, 529

\bibitem[{{Bovy} {et~al}\mbox{.}(2019){Bovy}, {Leung}, {Hunt}, {Mackereth}, {Garc{\'\i}a-Hern{\'a}ndez}, \& {Roman-Lopes}}]{Bovy+2019}
{Bovy} J., {Leung} H.~W., {Hunt} J. A.~S., {Mackereth} J.~T., {Garc{\'\i}a-Hern{\'a}ndez} D.~A., {Roman-Lopes} A., 2019, \mnras, 490, 4740

\bibitem[{{Chemin}, {Renaud} \& {Soubiran}(2015){Chemin}, {Renaud}, \& {Soubiran}}]{Chemin+2015}
{Chemin} L., {Renaud} F., {Soubiran} C., 2015, \aap, 578, A14

\bibitem[{{Courteau} {et~al}\mbox{.}(2014){Courteau}, {Cappellari}, {de Jong}, {Dutton}, {Emsellem}, {Hoekstra}, {Koopmans}, {Mamon}, {Maraston}, {Treu}, \& {Widrow}}]{Courteau+2014}
{Courteau} S. {et~al.}, 2014, Reviews of Modern Physics, 86, 47

\bibitem[{{Davis} {et~al}\mbox{.}(2026){Davis}, {Tepper-Garc{\'\i}a}, {McClure-Griffiths}, {Bland-Hawthorn}, \& {Agertz}}]{Davis+2026}
{Davis} H., {Tepper-Garc{\'\i}a} T., {McClure-Griffiths} N., {Bland-Hawthorn} J., {Agertz} O., 2026, \mnras, 547, staf2166

\bibitem[{{de Salas} \& {Widmark}(2021)}]{deSalasWidmark2021}
{de Salas} P.~F., {Widmark} A., 2021, Reports on Progress in Physics, 84, 104901

\bibitem[{{Drimmel} {et~al}\mbox{.}(2025){Drimmel}, {Khanna}, {Poggio}, \& {Skowron}}]{Drimmel+2025}
{Drimmel} R., {Khanna} S., {Poggio} E., {Skowron} D.~M., 2025, \aap, 698, A230

\bibitem[{{Efstathiou}, {Lake} \& {Negroponte}(1982){Efstathiou}, {Lake}, \& {Negroponte}}]{Efstathiou+1982}
{Efstathiou} G., {Lake} G., {Negroponte} J., 1982, \mnras, 199, 1069

\bibitem[{{Eilers} {et~al}\mbox{.}(2020){Eilers}, {Hogg}, {Rix}, {Frankel}, {Hunt}, {Fouvry}, \& {Buck}}]{Eilers+2020}
{Eilers} A.-C., {Hogg} D.~W., {Rix} H.-W., {Frankel} N., {Hunt} J. A.~S., {Fouvry} J.-B., {Buck} T., 2020, \apj, 900, 186

\bibitem[{{Eilers} {et~al}\mbox{.}(2019){Eilers}, {Hogg}, {Rix}, \& {Ness}}]{Eilers+2019}
{Eilers} A.-C., {Hogg} D.~W., {Rix} H.-W., {Ness} M.~K., 2019, \apj, 871, 120

\bibitem[{{Feng} {et~al}\mbox{.}(2026){Feng}, {Huang}, {Zhang}, \& {Liu}}]{Feng+2026}
{Feng} Q., {Huang} Y., {Zhang} H., {Liu} J., 2026, \mnras, 546, stag011

\bibitem[{{Fragkoudi} {et~al}\mbox{.}(2019){Fragkoudi}, {Katz}, {Trick}, {White}, {Di Matteo}, {Sormani}, {Khoperskov}, {Haywood}, {Hall{\'e}}, \& {G{\'o}mez}}]{Fragkoudi+2019}
{Fragkoudi} F. {et~al.}, 2019, \mnras, 488, 3324

\bibitem[{{Fujii} {et~al}\mbox{.}(2011){Fujii}, {Baba}, {Saitoh}, {Makino}, {Kokubo}, \& {Wada}}]{Fujii+2011}
{Fujii} M.~S., {Baba} J., {Saitoh} T.~R., {Makino} J., {Kokubo} E., {Wada} K., 2011, \apj, 730, 109

\bibitem[{{Fujii} {et~al}\mbox{.}(2018){Fujii}, {B{\'e}dorf}, {Baba}, \& {Portegies Zwart}}]{Fujii+2018}
{Fujii} M.~S., {B{\'e}dorf} J., {Baba} J., {Portegies Zwart} S., 2018, \mnras, 477, 1451

\bibitem[{{Fujii} {et~al}\mbox{.}(2019){Fujii}, {B{\'e}dorf}, {Baba}, \& {Portegies Zwart}}]{Fujii+2019}
{Fujii} M.~S., {B{\'e}dorf} J., {Baba} J., {Portegies Zwart} S., 2019, \mnras, 482, 1983

\bibitem[{{Funakoshi} {et~al}\mbox{.}(2024){Funakoshi}, {Matsunaga}, {Kawata}, {Baba}, {Taniguchi}, \& {Fujii}}]{Funakoshi+2024}
{Funakoshi} N., {Matsunaga} N., {Kawata} D., {Baba} J., {Taniguchi} D., {Fujii} M., 2024, \mnras

\bibitem[{{Gaia Collaboration} {et~al}\mbox{.}(2018){Gaia Collaboration}, {Katz}, {Antoja}, {Romero-G{\'o}mez}, {Drimmel}, {Reyl{\'e}}, {Seabroke}, {Soubiran}, {Babusiaux}, {Di Matteo}, \& et~al.}]{Gaia-Katz+2018}
{Gaia Collaboration} {et~al.}, 2018, \aap, 616, A11

\bibitem[{{Gaia Collaboration} {et~al}\mbox{.}(2016){Gaia Collaboration}, {Prusti}, {de Bruijne}, {Brown}, {Vallenari}, {Babusiaux}, {Bailer-Jones}, {Bastian}, {Biermann}, {Evans}, {Eyer}, {Jansen}, {Jordi}, {Klioner}, {Lammers}, {Lindegren}, {Luri}, {Mignard}, {Milligan}, {Panem}, {Poinsignon}, {Pourbaix}, {Randich}, {Sarri}, {Sartoretti}, {Siddiqui}, {Soubiran}, {Valette}, {van Leeuwen}, {Walton}, {Aerts}, {Arenou}, {Cropper}, {Drimmel}, {H{\o}g}, {Katz}, {Lattanzi}, {O'Mullane}, {Grebel}, {Holland}, {Huc}, {Passot}, {Bramante}, {Cacciari}, {Casta{\~n}eda}, {Chaoul}, {Cheek}, {De Angeli}, {Fabricius}, {Guerra}, {Hern{\'a}ndez}, {Jean-Antoine-Piccolo}, {Masana}, {Messineo}, {Mowlavi}, {Nienartowicz}, {Ord{\'o}{\~n}ez-Blanco}, {Panuzzo}, {Portell}, {Richards}, {Riello}, {Seabroke}, {Tanga}, {Th{\'e}venin}, {Torra}, {Els}, {Gracia-Abril}, {Comoretto}, {Garcia-Reinaldos}, {Lock}, {Mercier}, {Altmann}, {Andrae}, {Astraatmadja}, {Bellas-Velidis}, {Benson}, {Berthier}, {Blomme}, {Busso}, {Carry}, {Cellino},
  {Clementini}, {Cowell}, {Creevey}, {Cuypers}, {Davidson}, {De Ridder}, {de Torres}, {Delchambre}, {Dell'Oro}, {Ducourant}, {Fr{\'e}mat}, {Garc{\'\i}a-Torres}, {Gosset}, {Halbwachs}, {Hambly}, {Harrison}, {Hauser}, {Hestroffer}, {Hodgkin}, {Huckle}, {Hutton}, {Jasniewicz}, {Jordan}, {Kontizas}, {Korn}, {Lanzafame}, {Manteiga}, {Moitinho}, {Muinonen}, {Osinde}, {Pancino}, {Pauwels}, {Petit}, {Recio-Blanco}, {Robin}, {Sarro}, {Siopis}, {Smith}, {Smith}, {Sozzetti}, {Thuillot}, {van Reeven}, {Viala}, {Abbas}, {Abreu Aramburu}, {Accart}, {Aguado}, {Allan}, {Allasia}, {Altavilla}, {{\'A}lvarez}, {Alves}, {Anderson}, {Andrei}, {Anglada Varela}, {Antiche}, {Antoja}, {Ant{\'o}n}, {Arcay}, {Atzei}, {Ayache}, {Bach}, {Baker}, {Balaguer-N{\'u}{\~n}ez}, {Barache}, {Barata}, {Barbier}, {Barblan}, {Baroni}, {Barrado y Navascu{\'e}s}, {Barros}, {Barstow}, {Becciani}, {Bellazzini}, {Bellei}, {Bello Garc{\'\i}a}, {Belokurov}, {Bendjoya}, {Berihuete}, {Bianchi}, {Bienaym{\'e}}, {Billebaud}, {Blagorodnova}, {Blanco-Cuaresma},
  {Boch}, {Bombrun}, {Borrachero}, {Bouquillon}, {Bourda}, {Bouy}, {Bragaglia}, {Breddels}, {Brouillet}, {Br{\"u}semeister}, {Bucciarelli}, {Budnik}, {Burgess}, {Burgon}, {Burlacu}, {Busonero}, {Buzzi}, {Caffau}, {Cambras}, {Campbell}, {Cancelliere}, {Cantat-Gaudin}, {Carlucci}, {Carrasco}, {Castellani}, {Charlot}, {Charnas}, {Charvet}, {Chassat}, {Chiavassa}, {Clotet}, {Cocozza}, {Collins}, {Collins}, {Costigan}, {Crifo}, {Cross}, {Crosta}, {Crowley}, {Dafonte}, {Damerdji}, {Dapergolas}, {David}, {David}, {De Cat}, {de Felice}, {de Laverny}, {De Luise}, {De March}, {de Martino}, {de Souza}, {Debosscher}, {del Pozo}, {Delbo}, {Delgado}, {Delgado}, {di Marco}, {Di Matteo}, {Diakite}, {Distefano}, {Dolding}, {Dos Anjos}, {Drazinos}, {Dur{\'a}n}, {Dzigan}, {Ecale}, {Edvardsson}, {Enke}, {Erdmann}, {Escolar}, {Espina}, {Evans}, {Eynard Bontemps}, {Fabre}, {Fabrizio}, {Faigler}, {Falc{\~a}o}, {Farr{\`a}s Casas}, {Faye}, {Federici}, {Fedorets}, {Fern{\'a}ndez-Hern{\'a}ndez}, {Fernique}, {Fienga}, {Figueras},
  {Filippi}, {Findeisen}, {Fonti}, {Fouesneau}, {Fraile}, {Fraser}, {Fuchs}, {Furnell}, {Gai}, {Galleti}, {Galluccio}, {Garabato}, {Garc{\'\i}a-Sedano}, {Gar{\'e}}, {Garofalo}, {Garralda}, {Gavras}, {Gerssen}, {Geyer}, {Gilmore}, {Girona}, {Giuffrida}, {Gomes}, {Gonz{\'a}lez-Marcos}, {Gonz{\'a}lez-N{\'u}{\~n}ez}, {Gonz{\'a}lez-Vidal}, {Granvik}, {Guerrier}, {Guillout}, {Guiraud}, {G{\'u}rpide}, {Guti{\'e}rrez-S{\'a}nchez}, {Guy}, {Haigron}, {Hatzidimitriou}, {Haywood}, {Heiter}, {Helmi}, {Hobbs}, {Hofmann}, {Holl}, {Holland }, {Hunt}, {Hypki}, {Icardi}, {Irwin}, {Jevardat de Fombelle}, {Jofr{\'e}}, {Jonker}, {Jorissen}, {Julbe}, {Karampelas}, {Kochoska}, {Kohley}, {Kolenberg}, {Kontizas}, {Koposov}, {Kordopatis}, {Koubsky}, {Kowalczyk}, {Krone-Martins}, {Kudryashova}, {Kull}, {Bachchan}, {Lacoste-Seris}, {Lanza}, {Lavigne}, {Le Poncin-Lafitte}, {Lebreton}, {Lebzelter}, {Leccia}, {Leclerc}, {Lecoeur-Taibi}, {Lemaitre}, {Lenhardt}, {Leroux}, {Liao}, {Licata}, {Lindstr{\o}m}, {Lister}, {Livanou}, {Lobel},
  {L{\"o}ffler}, {L{\'o}pez}, {Lopez-Lozano}, {Lorenz}, {Loureiro}, {MacDonald}, {Magalh{\~a}es Fernandes}, {Managau}, {Mann}, {Mantelet}, {Marchal}, {Marchant}, {Marconi}, {Marie}, {Marinoni}, {Marrese}, {Marschalk{\'o}}, {Marshall}, {Mart{\'\i}n-Fleitas}, {Martino}, {Mary}, {Matijevi{\v{c}}}, {Mazeh}, {McMillan}, {Messina}, {Mestre}, {Michalik}, {Millar}, {Miranda}, {Molina}, {Molinaro}, {Molinaro}, {Moln{\'a}r}, {Moniez}, {Montegriffo}, {Monteiro}, {Mor}, {Mora}, {Morbidelli}, {Morel}, {Morgenthaler}, {Morley}, {Morris}, {Mulone}, {Muraveva}, {Musella}, {Narbonne}, {Nelemans}, {Nicastro}, {Noval}, {Ord{\'e}novic}, {Ordieres-Mer{\'e}}, {Osborne}, {Pagani}, {Pagano}, {Pailler}, {Palacin}, {Palaversa}, {Parsons}, {Paulsen}, {Pecoraro}, {Pedrosa}, {Pentik{\"a}inen}, {Pereira}, {Pichon}, {Piersimoni}, {Pineau}, {Plachy}, {Plum}, {Poujoulet}, {Pr{\v{s}}a}, {Pulone}, {Ragaini}, {Rago}, {Rambaux}, {Ramos-Lerate}, {Ranalli}, {Rauw}, {Read}, {Regibo}, {Renk}, {Reyl{\'e}}, {Ribeiro}, {Rimoldini}, {Ripepi}, {Riva},
  {Rixon}, {Roelens}, {Romero-G{\'o}mez}, {Rowell}, {Royer}, {Rudolph}, {Ruiz-Dern}, {Sadowski}, {Sagrist{\`a} Sell{\'e}s}, {Sahlmann}, {Salgado}, {Salguero}, {Sarasso}, {Savietto}, {Schnorhk}, {Schultheis}, {Sciacca}, {Segol}, {Segovia}, {Segransan}, {Serpell}, {Shih}, {Smareglia}, {Smart}, {Smith}, {Solano}, {Solitro}, {Sordo}, {Soria Nieto}, {Souchay}, {Spagna}, {Spoto}, {Stampa}, {Steele}, {Steidelm{\"u}ller}, {Stephenson}, {Stoev}, {Suess}, {S{\"u}veges}, {Surdej}, {Szabados}, {Szegedi-Elek}, {Tapiador}, {Taris}, {Tauran}, {Taylor}, {Teixeira}, {Terrett}, {Tingley}, {Trager}, {Turon}, {Ulla}, {Utrilla}, {Valentini}, {van Elteren}, {Van Hemelryck}, {van Leeuwen}, {Varadi}, {Vecchiato}, {Veljanoski}, {Via}, {Vicente}, {Vogt}, {Voss}, {Votruba}, {Voutsinas}, {Walmsley}, {Weiler}, {Weingrill}, {Werner}, {Wevers}, {Whitehead}, {Wyrzykowski}, {Yoldas}, {{\v{Z}}erjal}, {Zucker}, {Zurbach}, {Zwitter}, {Alecu}, {Allen}, {Allende Prieto}, {Amorim}, {Anglada-Escud{\'e}}, {Arsenijevic}, {Azaz}, {Balm}, {Beck},
  {Bernstein}, {Bigot}, {Bijaoui}, {Blasco}, {Bonfigli}, {Bono}, {Boudreault}, {Bressan}, {Brown}, {Brunet}, {Bunclark}, {Buonanno}, {Butkevich}, {Carret}, {Carrion}, {Chemin}, {Ch{\'e}reau}, {Corcione}, {Darmigny}, {de Boer}, {de Teodoro}, {de Zeeuw}, {Delle Luche}, {Domingues}, {Dubath}, {Fodor}, {Fr{\'e}zouls}, {Fries}, {Fustes}, {Fyfe}, {Gallardo}, {Gallegos}, {Gardiol}, {Gebran}, {Gomboc}, {G{\'o}mez}, {Grux}, {Gueguen}, {Heyrovsky}, {Hoar}, {Iannicola}, {Isasi Parache}, {Janotto}, {Joliet}, {Jonckheere}, {Keil}, {Kim}, {Klagyivik}, {Klar}, {Knude}, {Kochukhov}, {Kolka}, {Kos}, {Kutka}, {Lainey}, {LeBouquin}, {Liu}, {Loreggia}, {Makarov}, {Marseille}, {Martayan}, {Martinez-Rubi}, {Massart}, {Meynadier}, {Mignot}, {Munari}, {Nguyen}, {Nordlander}, {Ocvirk}, {O'Flaherty}, {Olias Sanz}, {Ortiz}, {Osorio}, {Oszkiewicz}, {Ouzounis}, {Palmer}, {Park}, {Pasquato}, {Peltzer}, {Peralta}, {P{\'e}turaud}, {Pieniluoma}, {Pigozzi}, {Poels}, {Prat}, {Prod'homme}, {Raison}, {Rebordao}, {Risquez}, {Rocca-Volmerange},
  {Rosen}, {Ruiz-Fuertes}, {Russo}, {Sembay}, {Serraller Vizcaino}, {Short}, {Siebert}, {Silva}, {Sinachopoulos}, {Slezak}, {Soffel}, {Sosnowska}, {Strai{\v{z}}ys}, {ter Linden}, {Terrell}, {Theil}, {Tiede}, {Troisi}, {Tsalmantza}, {Tur}, {Vaccari}, {Vachier}, {Valles}, {Van Hamme}, {Veltz}, {Virtanen}, {Wallut}, {Wichmann}, {Wilkinson}, {Ziaeepour}, \& {Zschocke}}]{GaiaCollaboration2016}
{Gaia Collaboration} {et~al.}, 2016, \aap, 595, A1

\bibitem[{{Grand}, {Kawata} \& {Cropper}(2012){Grand}, {Kawata}, \& {Cropper}}]{Grand+2012b}
{Grand} R.~J.~J., {Kawata} D., {Cropper} M., 2012, MNRAS, 426, 167

\bibitem[{{GRAVITY Collaboration} {et~al}\mbox{.}(2021){GRAVITY Collaboration}, {Abuter}, {Amorim}, {Baub{\"o}ck}, {Berger}, {Bonnet}, {Brandner}, {Cl{\'e}net}, {Davies}, {de Zeeuw}, {Dexter}, {Dallilar}, {Drescher}, {Eckart}, {Eisenhauer}, {F{\"o}rster Schreiber}, {Garcia}, {Gao}, {Gendron}, {Genzel}, {Gillessen}, {Habibi}, {Haubois}, {Hei{\ss}el}, {Henning}, {Hippler}, {Horrobin}, {Jim{\'e}nez-Rosales}, {Jochum}, {Jocou}, {Kaufer}, {Kervella}, {Lacour}, {Lapeyr{\`e}re}, {Le Bouquin}, {L{\'e}na}, {Lutz}, {Nowak}, {Ott}, {Paumard}, {Perraut}, {Perrin}, {Pfuhl}, {Rabien}, {Rodr{\'\i}guez-Coira}, {Shangguan}, {Shimizu}, {Scheithauer}, {Stadler}, {Straub}, {Straubmeier}, {Sturm}, {Tacconi}, {Vincent}, {von Fellenberg}, {Waisberg}, {Widmann}, {Wieprecht}, {Wiezorrek}, {Woillez}, {Yazici}, {Young}, \& {Zins}}]{GravityCollaboration+2021}
{GRAVITY Collaboration} {et~al.}, 2021, \aap, 647, A59

\bibitem[{{Hilmi} {et~al}\mbox{.}(2020){Hilmi}, {Minchev}, {Buck}, {Martig}, {Quillen}, {Monari}, {Famaey}, {de Jong}, {Laporte}, {Read}, {Sand ers}, {Steinmetz}, \& {Wegg}}]{Hilmi+2020}
{Hilmi} T. {et~al.}, 2020, \mnras, 497, 933

\bibitem[{{Hou} \& {Han}(2014)}]{HouHan2014}
{Hou} L.~G., {Han} J.~L., 2014, \aap, 569, A125

\bibitem[{{Hunt} {et~al}\mbox{.}(2018){Hunt}, {Hong}, {Bovy}, {Kawata}, \& {Grand}}]{Hunt+2018}
{Hunt} J. A.~S., {Hong} J., {Bovy} J., {Kawata} D., {Grand} R. J.~J., 2018, \mnras, 481, 3794

\bibitem[{{Hunt} {et~al}\mbox{.}(2021){Hunt}, {Stelea}, {Johnston}, {Gandhi}, {Laporte}, \& {B{\'e}dorf}}]{Hunt+2021}
{Hunt} J. A.~S., {Stelea} I.~A., {Johnston} K.~V., {Gandhi} S.~S., {Laporte} C. F.~P., {B{\'e}dorf} J., 2021, \mnras, 508, 1459

\bibitem[{{Hunt} \& {Vasiliev}(2025)}]{HuntVasiliev2025}
{Hunt} J. A.~S., {Vasiliev} E., 2025, \nar, 100, 101721

\bibitem[{{Jiao} {et~al}\mbox{.}(2023){Jiao}, {Hammer}, {Wang}, {Wang}, {Amram}, {Chemin}, \& {Yang}}]{Jiao+2023}
{Jiao} Y., {Hammer} F., {Wang} H., {Wang} J., {Amram} P., {Chemin} L., {Yang} Y., 2023, \aap, 678, A208

\bibitem[{{Kawata} {et~al}\mbox{.}(2018){Kawata}, {Baba}, {Ciuc{\v{a}}}, {Cropper}, {Grand}, {Hunt}, \& {Seabroke}}]{Kawata+2018}
{Kawata} D., {Baba} J., {Ciuc{\v{a}}} I., {Cropper} M., {Grand} R. J.~J., {Hunt} J. A.~S., {Seabroke} G., 2018, \mnras, 479, L108

\bibitem[{{Kawata} {et~al}\mbox{.}(2019){Kawata}, {Bovy}, {Matsunaga}, \& {Baba}}]{Kawata+2019}
{Kawata} D., {Bovy} J., {Matsunaga} N., {Baba} J., 2019, \mnras, 482, 40

\bibitem[{{Koop} {et~al}\mbox{.}(2024){Koop}, {Antoja}, {Helmi}, {Callingham}, \& {Laporte}}]{Koop+2024}
{Koop} O., {Antoja} T., {Helmi} A., {Callingham} T.~M., {Laporte} C. F.~P., 2024, \aap, 692, A50

\bibitem[{{Laporte} {et~al}\mbox{.}(2019){Laporte}, {Minchev}, {Johnston}, \& {G{\'o}mez}}]{Laporte+2019}
{Laporte} C. F.~P., {Minchev} I., {Johnston} K.~V., {G{\'o}mez} F.~A., 2019, \mnras, 485, 3134

\bibitem[{{Leung} {et~al}\mbox{.}(2023){Leung}, {Bovy}, {Mackereth}, {Hunt}, {Lane}, \& {Wilson}}]{Leung+2023}
{Leung} H.~W., {Bovy} J., {Mackereth} J.~T., {Hunt} J. A.~S., {Lane} R.~R., {Wilson} J.~C., 2023, \mnras, 519, 948

\bibitem[{{Lin} {et~al}\mbox{.}(2022){Lin}, {Xu}, {Hou}, {Liu}, {Li}, {Hao}, {Li}, \& {Bian}}]{Lin+2022}
{Lin} Z., {Xu} Y., {Hou} L., {Liu} D., {Li} Y., {Hao} C., {Li} J., {Bian} S., 2022, \apj, 931, 72

\bibitem[{{Martinez-Medina}, {P{\'e}rez-Villegas} \& {Peimbert}(2022){Martinez-Medina}, {P{\'e}rez-Villegas}, \& {Peimbert}}]{Martinez-Medina+2022}
{Martinez-Medina} L., {P{\'e}rez-Villegas} A., {Peimbert} A., 2022, \mnras, 512, 1574

\bibitem[{{Martinez-Medina} {et~al}\mbox{.}(2019){Martinez-Medina}, {Pichardo}, {Peimbert}, \& {Valenzuela}}]{Martinez-Medina+2019}
{Martinez-Medina} L., {Pichardo} B., {Peimbert} A., {Valenzuela} O., 2019, \mnras, 485, L104

\bibitem[{{McMillan}(2017)}]{McMillan2017}
{McMillan} P.~J., 2017, \mnras, 465, 76

\bibitem[{{Miyachi} {et~al}\mbox{.}(2019){Miyachi}, {Sakai}, {Kawata}, {Baba}, {Honma}, {Matsunaga}, \& {Fujisawa}}]{Miyachi+2019}
{Miyachi} Y., {Sakai} N., {Kawata} D., {Baba} J., {Honma} M., {Matsunaga} N., {Fujisawa} K., 2019, \apj, 882, 48

\bibitem[{{Monari} {et~al}\mbox{.}(2016){Monari}, {Famaey}, {Siebert}, {Grand }, {Kawata}, \& {Boily}}]{Monari+2016b}
{Monari} G., {Famaey} B., {Siebert} A., {Grand } R. J.~J., {Kawata} D., {Boily} C., 2016, \mnras, 461, 3835

\bibitem[{{Ou} {et~al}\mbox{.}(2024){Ou}, {Eilers}, {Necib}, \& {Frebel}}]{Ou+2024}
{Ou} X., {Eilers} A.-C., {Necib} L., {Frebel} A., 2024, \mnras, 528, 693

\bibitem[{{P{\~o}der} {et~al}\mbox{.}(2023){P{\~o}der}, {Benito}, {Pata}, {Kipper}, {Ramler}, {H{\"u}tsi}, {Kolka}, \& {Thomas}}]{Poder+2023}
{P{\~o}der} S., {Benito} M., {Pata} J., {Kipper} R., {Ramler} H., {H{\"u}tsi} G., {Kolka} I., {Thomas} G.~F., 2023, \aap, 676, A134

\bibitem[{{Perryman}(2026)}]{Perryman2026}
{Perryman} M., 2026, \physrep, 1150, 1

\bibitem[{{Poggio} {et~al}\mbox{.}(2021){Poggio}, {Drimmel}, {Cantat-Gaudin}, {Ramos}, {Ripepi}, {Zari}, {Andrae}, {Blomme}, {Chemin}, {Clementini}, {Figueras}, {Fouesneau}, {Fr{\'e}mat}, {Lobel}, {Marshall}, {Muraveva}, \& {Romero-G{\'o}mez}}]{Poggio+2021}
{Poggio} E. {et~al.}, 2021, \aap, 651, A104

\bibitem[{{Portail} {et~al}\mbox{.}(2017){Portail}, {Gerhard}, {Wegg}, \& {Ness}}]{Portail+2017}
{Portail} M., {Gerhard} O., {Wegg} C., {Ness} M., 2017, \mnras, 465, 1621

\bibitem[{{Ramos}, {Antoja} \& {Figueras}(2018){Ramos}, {Antoja}, \& {Figueras}}]{Ramos+2018}
{Ramos} P., {Antoja} T., {Figueras} F., 2018, \aap, 619, A72

\bibitem[{{Reid} {et~al}\mbox{.}(2019){Reid}, {Menten}, {Brunthaler}, {Zheng}, {Dame}, {Xu}, {Li}, {Sakai}, {Wu}, {Immer}, {Zhang}, {Sanna}, {Moscadelli}, {Rygl}, {Bartkiewicz}, {Hu}, {Quiroga-Nu{\~n}ez}, \& {van Langevelde}}]{Reid+2019}
{Reid} M.~J. {et~al.}, 2019, \apj, 885, 131

\bibitem[{{Saitoh}(2017)}]{Saitoh2017}
{Saitoh} T.~R., 2017, \aj, 153, 85

\bibitem[{{Saitoh} {et~al}\mbox{.}(2008){Saitoh}, {Daisaka}, {Kokubo}, {Makino}, {Okamoto}, {Tomisaka}, {Wada}, \& {Yoshida}}]{Saitoh+2008}
{Saitoh} T.~R., {Daisaka} H., {Kokubo} E., {Makino} J., {Okamoto} T., {Tomisaka} K., {Wada} K., {Yoshida} N., 2008, \pasj, 60, 667

\bibitem[{{Saitoh} \& {Makino}(2013)}]{SaitohMakino2013}
{Saitoh} T.~R., {Makino} J., 2013, \apj, 768, 44

\bibitem[{{Sellwood} \& {Sparke}(1988)}]{SellwoodSparke1988}
{Sellwood} J.~A., {Sparke} L.~S., 1988, MNRAS, 231, 25P

\bibitem[{{Sharma} {et~al}\mbox{.}(2021){Sharma}, {Hayden}, {Bland-Hawthorn}, {Stello}, {Buder}, {Zinn}, {Kallinger}, {Asplund}, {De Silva}, {D'Orazi}, {Freeman}, {Kos}, {Lewis}, {Lin}, {Lind}, {Martell}, {Simpson}, {Wittenmyer}, {Zucker}, {Zwitter}, {Chen}, {Cotar}, {Esdaile}, {Hon}, {Horner}, {Huber}, {Kafle}, {Khanna}, {Ting}, {Nataf}, {Nordlander}, {Saadon}, {Tepper-Garcia}, {Tinney}, {Traven}, {Watson}, {Wright}, \& {Wyse}}]{Sharma+2021a}
{Sharma} S. {et~al.}, 2021, \mnras, 506, 1761

\bibitem[{{van Albada} {et~al}\mbox{.}(1985){van Albada}, {Bahcall}, {Begeman}, \& {Sancisi}}]{vanAlbada+1985}
{van Albada} T.~S., {Bahcall} J.~N., {Begeman} K., {Sancisi} R., 1985, \apj, 295, 305

\bibitem[{{Vasiliev}(2019)}]{Vasiliev2019}
{Vasiliev} E., 2019, \mnras, 482, 1525

\bibitem[{{Vislosky} {et~al}\mbox{.}(2024){Vislosky}, {Minchev}, {Khoperskov}, {Martig}, {Buck}, {Hilmi}, {Ratcliffe}, {Bland-Hawthorn}, {Quillen}, {Steinmetz}, \& {de Jong}}]{Vislosky+2024}
{Vislosky} E. {et~al.}, 2024, \mnras, 528, 3576

\bibitem[{{Wegg} \& {Gerhard}(2013)}]{WeggGerhard2013}
{Wegg} C., {Gerhard} O., 2013, \mnras, 435, 1874

\bibitem[{{Zhou} {et~al}\mbox{.}(2023){Zhou}, {Li}, {Huang}, \& {Zhang}}]{Zhou+2023}
{Zhou} Y., {Li} X., {Huang} Y., {Zhang} H., 2023, \apj, 946, 73

\end{thebibliography}

\end{document}